\documentclass[journal]{IEEEtran}

\usepackage{cite}
\usepackage{amsmath,amssymb,amsfonts}
\usepackage{algorithmic}
\usepackage{graphicx}
\usepackage{textcomp}
\usepackage{multirow}

\ifCLASSINFOpdf
\else
\fi
\begin{document}
%
\title{Identification of EEG Dynamics during Freezing of Gait and Voluntary Stopping in Patients with Parkinson's Disease }
%
%
%

\author{Zehong Cao $^\dagger$,~\IEEEmembership{Member,~IEEE,}
Alka Rachel John $^\dagger$,
Hsiang-Ting Chen,~\IEEEmembership{Senior Member,~IEEE,}
Kaylena Ehgoetz Martens,
Matthew Georgiades,
Moran Gilat,
Hung T. Nguyen, ~\IEEEmembership{Senior Member,~IEEE,}
Simon J. G. Lewis *,
and~Chin-Teng Lin *,~\IEEEmembership{Fellow,~IEEE}

\thanks{Zehong Cao was with Australian Artificial Intelligence Institute, Faculty of Engineering and Information Technology, University of Technology Sydney, NSW, Australia. He is now with the School of Information and Communication Technology, University of Tasmania, TAS, Australia.}
\thanks{Alka Rachel John is with the Australian Artificial Intelligence Institute, Faculty of Engineering and Information Technology, University of Technology Sydney, NSW, Australia.}
\thanks{Hsiang-Ting Chen was with Australian Artificial Intelligence Institute, Faculty of Engineering and Information Technology, University of Technology Sydney, NSW, Australia. He is now with the School of Computer Science, University of Adelaide, SA, Australia.}
\thanks{Kaylena Ehgoetz Martens was with the Parkinson's Disease Research Clinic, Brain and Mind Centre, University of Sydney, NSW, Australia. She is now with the Department of Kinesiology, University of Waterloo, Ontario, Canada.}
\thanks{Matthew Georgiades is with the Parkinson's Disease Research Clinic, Brain and Mind Centre, University of Sydney, NSW, Australia.}
\thanks{Moran Gilat was with the Parkinson's Disease Research Clinic, Brain and Mind Centre, University of Sydney, NSW, Australia. He is now with Department of Rehabilitation Sciences, KU Leuven, Leuven, Belgium.}
\thanks{Hung T. Nguyen is with the Faculty of Science, Engineering and Technology, Swinburne University of Technology, VIC, Australia.}
\thanks{Simon J. G. Lewis is with the Parkinson's Disease Research Clinic, Brain and Mind Centre, University of Sydney, NSW, Australia. (* E-mail: profsimonlewis@gmail.com).}
\thanks{Chin-Teng Lin is with the Australian Artificial Intelligence Institute, Faculty of Engineering and Information Technology, University of Technology Sydney, NSW, Australia. (* E-mail: Chin-Teng.Lin@uts.edu.au).}

\thanks{$\dagger$ Equal contribution. * Corresponding authors.} 
\thanks{Manuscript received xx, 2021.}}

%
%

\markboth{Journal of \LaTeX\ Class Files,~Vol.~xx, No.~xx, 2021}%
{Cao \MakeLowercase{\textit{et al.}}: Identification of EEG Dynamics During Freezing of Gait and Voluntary Stopping in Patients with Parkinson's Disease}
%



\maketitle

\begin{abstract}
Mobility is severely impacted in patients with Parkinson's disease (PD), especially when they experience involuntary stopping from the freezing of gait (FOG). Understanding the neurophysiological difference between ``voluntary stopping" and ``involuntary stopping” caused by FOG is vital for the detection and potential intervention of FOG in the daily lives of patients. This study characterised the electroencephalographic (EEG) signature associated with FOG in contrast to voluntary stopping. The protocol consisted of a timed up-and-go (TUG) task and an additional TUG task with a voluntary stopping component, where participants reacted to verbal ``stop" and ``walk" instructions by voluntarily stopping or walking. Event-related spectral perturbation (ERSP) analysis was used to study the dynamics of the EEG spectra induced by different walking phases, which included normal walking, voluntary stopping and episodes of involuntary stopping (FOG), as well as the transition windows between normal walking and voluntary stopping or FOG. These results demonstrate for the first time that the EEG signal during the transition from walking to voluntary stopping is distinguishable from that of the transition to involuntary stopping caused by FOG. The EEG signature of voluntary stopping exhibits a significantly decreased power spectrum compared to that of FOG episodes, with distinctly different patterns in the delta and low-beta power in the central area. These findings suggest the possibility of a practical EEG-based treatment strategy that can accurately predict FOG episodes, excluding the potential confound of voluntary stopping.
\end{abstract}


%
\IEEEpeerreviewmaketitle

\section{Introduction}
%
%
%
%
\IEEEPARstart{F}{reezing} of gait (FOG) is a devastating symptom of Parkinson's disease (PD) in which patients suddenly feel as though their feet have become ``stuck to the ground” \cite{b1}. Approximately 80\% of patients with severe PD are affected by FOG episodes, which often precipitate falls, leading to a high morbidity and the urgent need for nursing home placement \cite{b2}.

The current pathophysiology underlying the freezing phenomenon is not well-understood \cite{b3}. Some previous studies, such as functional magnetic resonance imaging (MRI)-based work, have attempted to understand the pathophysiology of FOG episodes and have identified some distinct patterns associated with freezing \cite{b4},\cite{b5}. Additionally, several electroencephalography (EEG)-based studies have also attempted to establish the neurophysiological correlate of FOG in PD. Increased beta power was previously observed in the subthalamic nucleus (STN) of patients with FOG \cite{b6} and more recent work has shown a temporal relationship with an increase in pathological beta and theta rhythms in the STN recorded during the performance of a virtual reality gait paradigm \cite{b18}.

Previously the brain dynamics associated with FOG episodes during turning measured by using ambulatory EEG has revealed significant changes in the high-beta and theta power spectral densities across the occipital and parietal areas during FOG episodes with turning \cite{b7}. In addition, EEG dynamics have demonstrated great potential in identifying the onset of freezing in patients with PD \cite{b8}. Indeed, EEG features have been suggested to be useful in predicting the transition from normal walking to freezing by using a 5-s time window before the episode \cite{b7}, \cite{b9}.

The advances in EEG have made it an efficient tool for understanding not only FOG episodes but also other ordinary motor tasks related to movements, including walking. For example, two studies \cite{b10}, \cite{b11} have suggested specific roles for EEG activity within particular frequency bands in the completion of ongoing motor tasks. Specifically, this EEG activity includes beta activity during motor preparation, gamma activity during motor commission and gating \cite{b12}, \cite{b13}, and theta activity during the processing of conflict-related signals \cite{b14}, \cite{b15}. Additionally, event-related EEG potentials, particularly the EEG signals occurring before the initiation of an action, were found to be useful in identifying the intention to move \cite{b16,b17}.

Since gait dynamics alone cannot accurately predict freezing episodes or independently distinguish freezing from voluntary stopping, we hypothesise that EEG could prove useful for the reliable detection of FOG. Anticipating that the brain dynamics during the transition to freezing can be confidently discerned from those during the transition to voluntary stopping, ``real-time'' EEG may offer a novel therapeutic intervention for the prediction and alleviation of freezing episodes. Recently, researchers have demonstrated that pathological subthalamic nucleus activity associated with bouts of freezing was discernible from that of voluntary stopping when assessed using a virtual ``gait'' paradigm while lying down and navigating the virtual environment by using a set of foot pedals \cite{b18}. This research provides positive support, but little work has been done to identify how brain dynamics during freezing are distinct from those during voluntary stopping while walking using ambulatory EEG.

In our study, to discern the neurophysiological differences between freezing and voluntary stopping episodes, we conducted the experiment building upon the timed up-and-go (TUG) protocol \cite{b19}. The TUG protocol consists of a sequence of sit-to-stand, walking, turning, and stand-to-sit tasks, each of which can be affected by freezing, especially when performed as a sequence \cite{b20}. In our study, we further added a condition of ``voluntary stopping'', where the participants reacted to the verbal instruction to voluntarily ``stop" while performing the TUG task. We tracked the EEG dynamics of patients with PD before and during FOG episodes and during voluntary stopping. Contrasting the signature of freezing from that of voluntary stopping could significantly help pave the way towards more effective therapeutics that accurately predict FOG events while excluding potential false positives associated with voluntary stopping.

We hope that our findings will provide a potential avenue for therapeutic prediction and alleviation of freezing episodes in patients with PD and promote exploration of voluntary stopping in the gait cycle by characterising the EEG signatures, as the identification of these movement intentions could be very useful in motor rehabilitation processes.

\section{MATERIALS AND METHODS}

\subsection{Subjects}
In this study, seventeen (17) patients from the Parkinson's Disease Research Clinic at the Brain and Mind Centre, University of Sydney, were identified by using the score for item 3 of the self-reported FOG Questionnaire (FOGQ), which was further confirmed by specialist review. Consistent with our previous studies \cite{b7, b21}, all patients satisfied the UK Parkinson's Disease Society Brain Bank (UKPDSBB) criteria, had a Mini-Mental State Examination (MMSE) score $\ge$ 24 and were deemed unlikely to have dementia or major depression according to DSM-IV criteria by the judgement of neurologist Simon J. G. Lewis. This study was approved by The University of Sydney Human Research and Ethics Committee, and written informed consent was obtained from all the patients.

\subsection{Experimental design}

All patients underwent a structured series of video-recorded TUG tasks while in their practically defined off state (having withdrawn from PD medication overnight for a minimum of 12 hours). All TUG tasks started in a seated position on a chair, from which patients walked along the centre of a large open corridor. A target box (0.6 m x 0.6 m) located six metres (6 m) from the chair was marked on the floor with white tape, and turning movements were performed in this box. A TUG task involved a 180° or 540° turn within the box and a return to the starting chair. It was performed with counterbalanced turns to the patient's left or right side. Furthermore, the videos were independently reviewed, and FOG episodes were scored by two experienced clinical researchers.

\begin{figure}[!t]
\centerline{\includegraphics[width=\columnwidth]{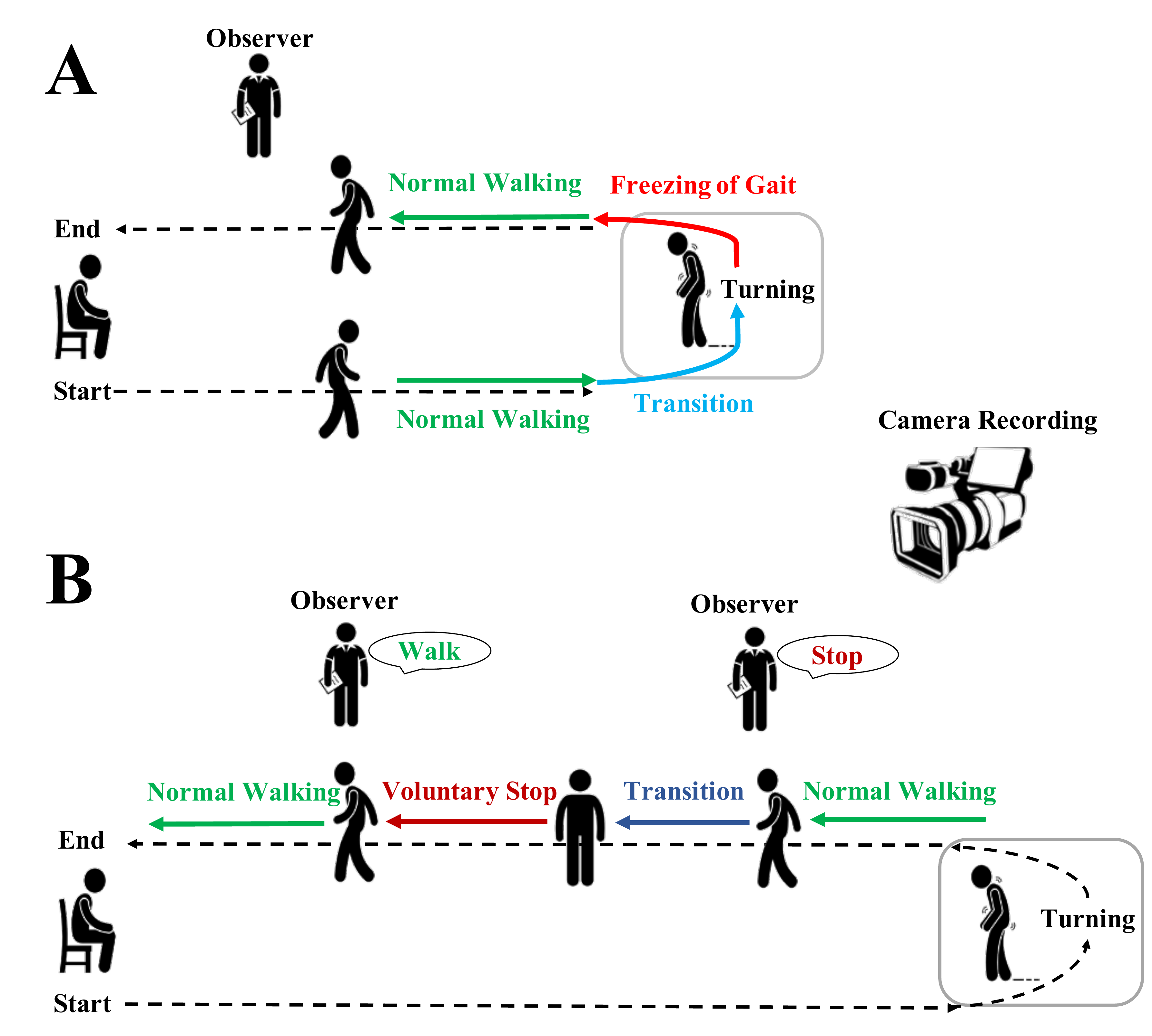}}
\caption{Experimental protocol of the TUG tasks. (A) The standard TUG task. (B) The TUG task with a voluntary stopping component with verbal instructions to ``stop'' and ``walk'' for assessing voluntary stopping.}
\label{fig1}
\end{figure}

Our experimental paradigm consisted of two types of TUG tasks: a standard TUG task to trigger FOG and a TUG task with a voluntary stopping component in which ``stop'' or ``walk'' verbal instructions were provided by the investigators to guide voluntary walking or stopping. For the standard TUG task (Fig.~\ref{fig1}A), patients performed a series of timed TUG tasks on a standardised course to evoke FOG events \cite{b20}. Each FOG trial consisted of three epochs. Specifically, we defined a ``normal walking'' epoch as a 2-s period in which the patient with PD walked normally with no cessation. In contrast, a ``transition'' epoch was the 2-s period before a freezing episode. Finally, a ``FOG'' episode was defined as an involuntary stop and was identified as the 2-s period from when a patient experienced an unwanted cessation in their normal stride.

For the TUG with a voluntary stopping component with verbal instructions to ``stop'' and ``walk'' (Fig.~\ref{fig1}B), a target box (0.6 m x 0.6 m) more than 10 m from the chair was marked on the floor with white tape, and turning movements were performed in this box. To promote voluntary stopping in the patients, the observer said the word ``stop'', and the patients were required to stop immediately. In the next 5-10 s, the observer said the word ``walk'', and the subjects were asked to start walking immediately. These voluntary stopping trials also consisted of three epochs. More specifically, we defined a ``normal walking'' epoch as a 2-s period in which the patient was walking normally with no cessation. A ``transition'' epoch was identified as the interval between giving a ``stop'' instruction and when the patient physically stopped walking. A ``voluntary stopping'' episode was identified as the 2-s epoch from when the patient physically stopped walking.

\subsection{Data recording and analysis}
Each patient wore a wearable 
BioSemi Active-Two system 
with 32 Ag-AgCl electrodes to record the EEG signals. The 32 channel electrodes (Fp1, Fp2, F7, F3, Fz, F4, F8, FT7, FC3, FCz, FC4, FT8, T7, C3, Cz, C4, T8, TP7, CP3, CPZ, CP4, TP8, P7, P3, Pz, P4, P8, O1, Oz, O2, A1, and A2) used the International 10–20 system locations to maintain a standard clinical setting, and the A1 and A2 electrodes were used as the reference channels. Before calibrating the electrodes, the patient's skin under the reference electrodes was washed with 70\% isopropyl alcohol. The EEG signals were recorded at a sample rate of 500 Hz with 16-bit quantisation.

\subsection{EEG pre-processing}
\begin{figure}[!t]
\centerline{\includegraphics[width=\columnwidth]{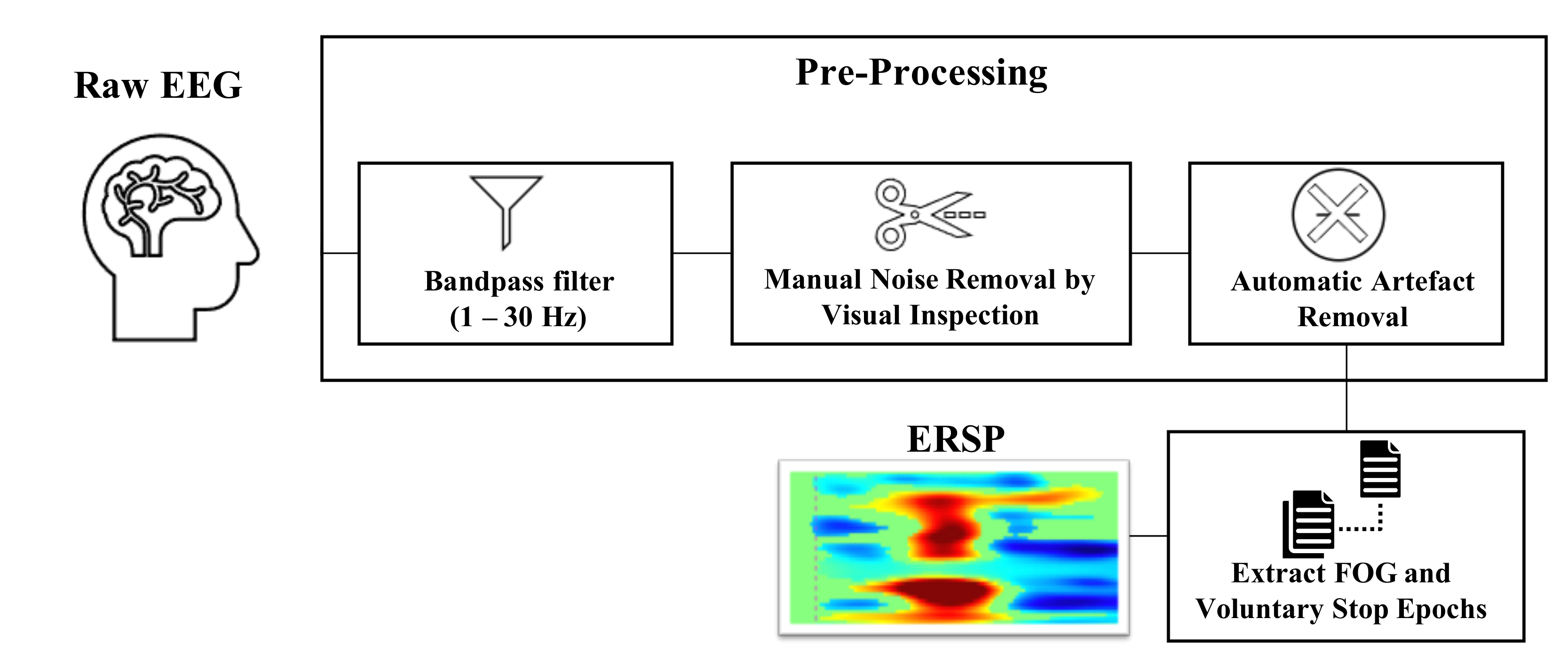}}
\caption{Diagram of the EEG pre-processing and analysis scheme.}
\label{fig2}
\end{figure}

The EEG data were analysed with the EEGLAB toolbox \cite{b23} in MATLAB R2016a. As shown in Fig. \ref{fig2}, the raw EEG signals were initially pre-processed by a 1-30-Hz bandpass filter to remove line noise, low-frequency noise and high-frequency components. Removing artefacts, such as eye movement and muscle activity, is crucial, as they can adversely affect further processing steps. For artefact rejection, eye movement contaminants in the EEG signals were identified by visual inspection and manually removed.

Additionally, artefacts were removed by the Automatic Artifact Removal (AAR) \cite{b24, b25} plug-in for EEGLAB, which integrates many state-of-the-art methods for automatic correction of ocular and muscular artefacts in the EEG signals. This toolbox uses blind source separation to decompose the EEG data into several spatial components before automatically removing artefact-related components. It then reconstructs the EEG signal by using the non-artefactual components. Afterwards, the FOG and voluntary stopping trials were extracted for further analyses, and a time-warping method was used to measure latencies.

In this study, a total of 178 FOG episodes and 54 voluntary stopping episodes were analysed. These 178 FOG episodes were extracted from both TUG tasks, of which 74 FOG episodes occurred during turning, while 104 occurred during the normal stride.

\subsection{EEG ERSP analysis}
The FOG and voluntary stopping epochs were extracted from the continuous EEG signals, and each epoch contained the sampled EEG data from -2000 ms to 4000 ms with the stimulus onset at 0 ms. To investigate brain dynamics during the FOG and voluntary stopping episodes and the subsequent motor responses, each epoch was separately transformed into a time-frequency representation by using an event-related spectral perturbation (ERSP) routine \cite{b26}, which is a time-frequency analysis to transform time-domain signals into the spectral-temporal domain to characterise the event-related frequency changes by using the 
short-time Fourier transform (STFT).

To compute the ERSP for each trial, spectra prior to the event onset are considered the baseline spectra for that trial. The mean of the baseline spectra (in dB) is subtracted from the spectral power after event onset to visualise the spectral ``perturbation'' from the baseline. The mean normal walking log power spectrum (in dB) of the optimal epochs, which was used as the reference (baseline) value, was subtracted from each estimated spectrum. The ERSP is derived by computing the power spectrum over a sliding latency window and then computing the average across all the trials. Therefore, for n trials, if $F_{k}(f, t)$ is the spectral estimate of trial $k$ at frequency $f$ and time $t$, then the ERSP is given by
\begin{equation} ERSP(f,t)=\frac{1}{n}*\sum |F_{k}(f,t)|. \label{eq} \end{equation}

In EEGLAB, $F_{k}(f, t)$ is calculated by using the STFT \cite{b27}. 
The relative power at a particular frequency and latency is indicated by the colour of the corresponding pixel in the ERSP image.

In this study, the mean power spectra, including the delta (1-3.5 Hz), theta (4-7.5 Hz), alpha (8-12.5 Hz), and beta (13-30 Hz) bands, were vertically stacked from normal walking and the transition to FOG or voluntary stopping. We considered four important electrodes (Fz – supplementary motor area, Cz – primary motor area, P4 – navigational movement area, and O1 – primary visual receiving area), as well as the left-hemisphere (F3, FC3, C3, CP3, and P3) and right-hemisphere (F4, FC4, C4, CP4, and P4) brain areas, to explore event-related band power dynamics that accompanied the transition of power changes before and after the FOG and voluntary stopping episodes (the x-axis is time, and the y-axis is the power spectrum).

\subsection{Statistical analysis}

The ERSPs for normal walking, the transition to FOG and the transition to voluntary stopping were compared using paired t-tests. Independent t-tests were performed to compare the ERSPs between the FOG and voluntary stopping episodes. The significance level was set at $\alpha < 0.05$. The statistical analysis was performed in MATLAB (R2016a).

\section{Results}
\subsection{Event-related band power dynamics during FOG in the TUG trials}
\begin{figure}[!t]
\centerline{\includegraphics[width=6.5cm]{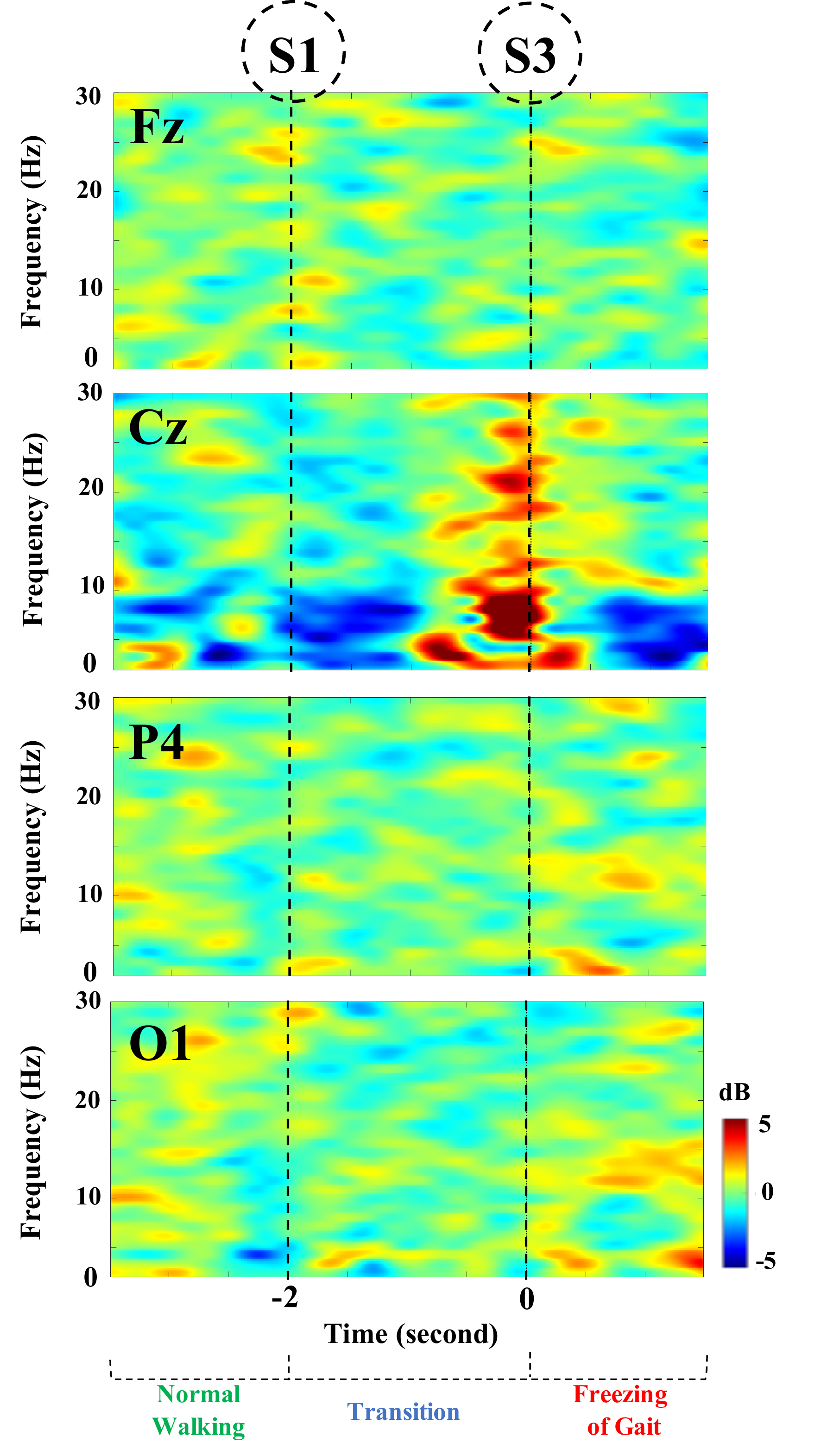}}
\caption{Event-related band power dynamics during FOG episodes while performing the TUG task. S1 denotes the timing from normal walking to the transition period, and S3 denotes the timing from the transition period to FOG. The ERSP changes from normal walking ($<$ -2 s), transition (-2 s to 0 s), to FOG ($>$ 0 s) at four electrodes (Fz, Cz, P4 and O1) .}
\label{fig3}
\end{figure}

The two-dimensional images in Fig.~\ref{fig3} show the ERSPs of the frontal, central, parietal, and occipital channels for patients with PD accompanied by FOG. As shown in Fig.~\ref{fig3}, the mean EEG power spectra before and after the onset of freezing at the Fz, Cz, P4 and O1 electrodes are presented. In particular, the ERSP at the Cz electrode showed a decreased EEG power spectrum during the transition period relative to normal walking and an increased EEG power spectrum in the transition period relative to FOG episodes.

\begin{figure}[!t]
\centerline{\includegraphics[width=\columnwidth]{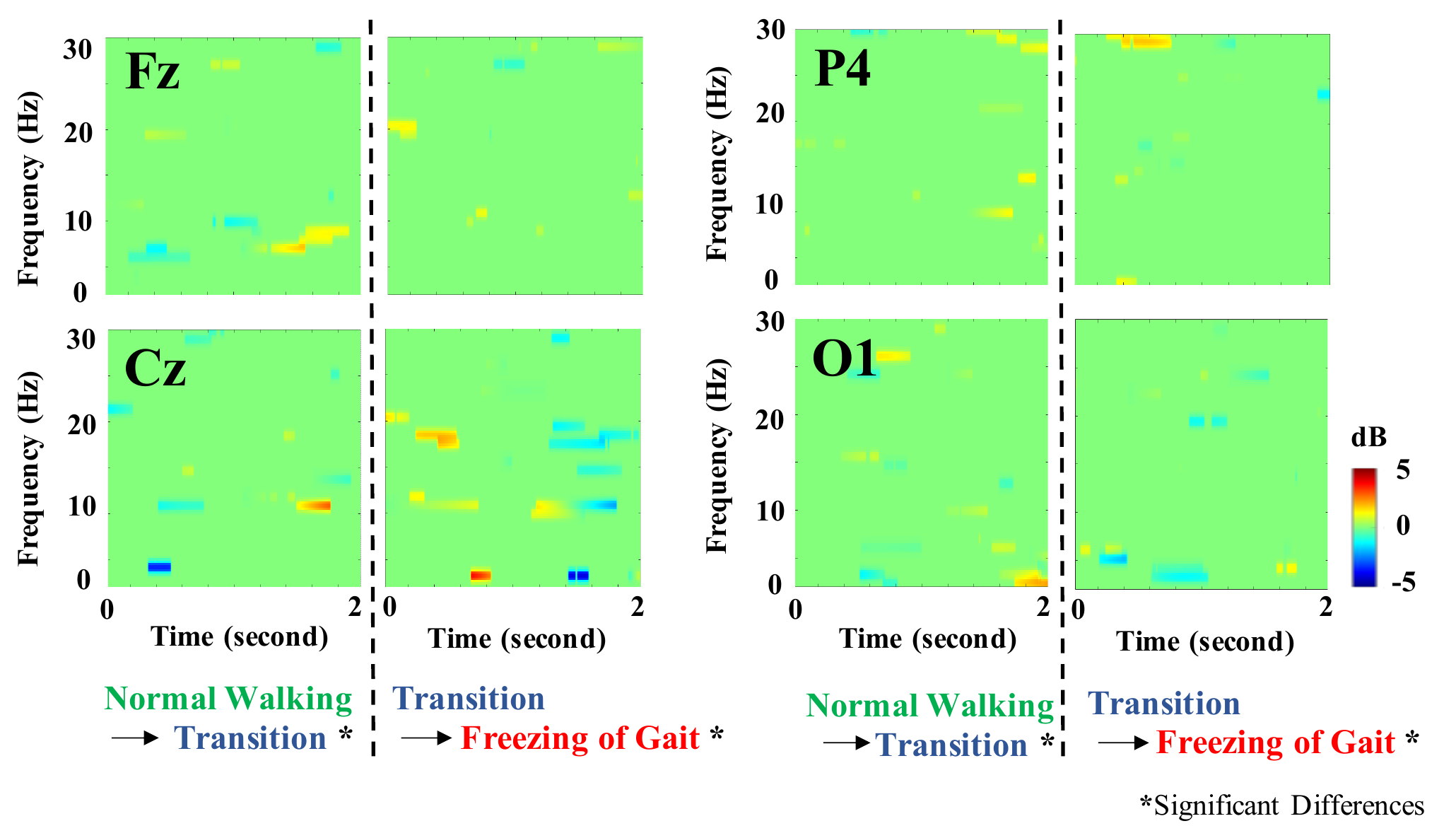}}
\caption{Comparisons of the ERSP differences between the normal walking and the transition period (left column) and the transition period and the FOG episode (right column) at four important electrodes (Fz, Cz, P4 and O1). }
\label{fig4}
\end{figure}

Furthermore, we compared the different periods during FOG episodes, as shown in Fig.~\ref{fig4}. Compared to those during normal walking, the EEG alpha power band and a portion of the beta power band in the transition period were significantly elevated at the Cz and O1 electrodes (p $<$ 0.05). The ERSP changes during the transition period and at the onset of FOG showed partial increases in the beta power band at the Fz and P4 electrodes, significantly lower theta and alpha power bands early in the time interval at the Fz, Cz, P4 and O1 electrodes, and partial increases in the EEG power spectrum late in the time interval at the P4 and O1 electrodes (p $<$ 0.05) during the transition period relative to the FOG episodes.

\subsection{Event-related band power dynamics between voluntary stopping and FOG episodes during the TUG trials}

\begin{figure*}[!ht]
\center
    \includegraphics[width=\textwidth]{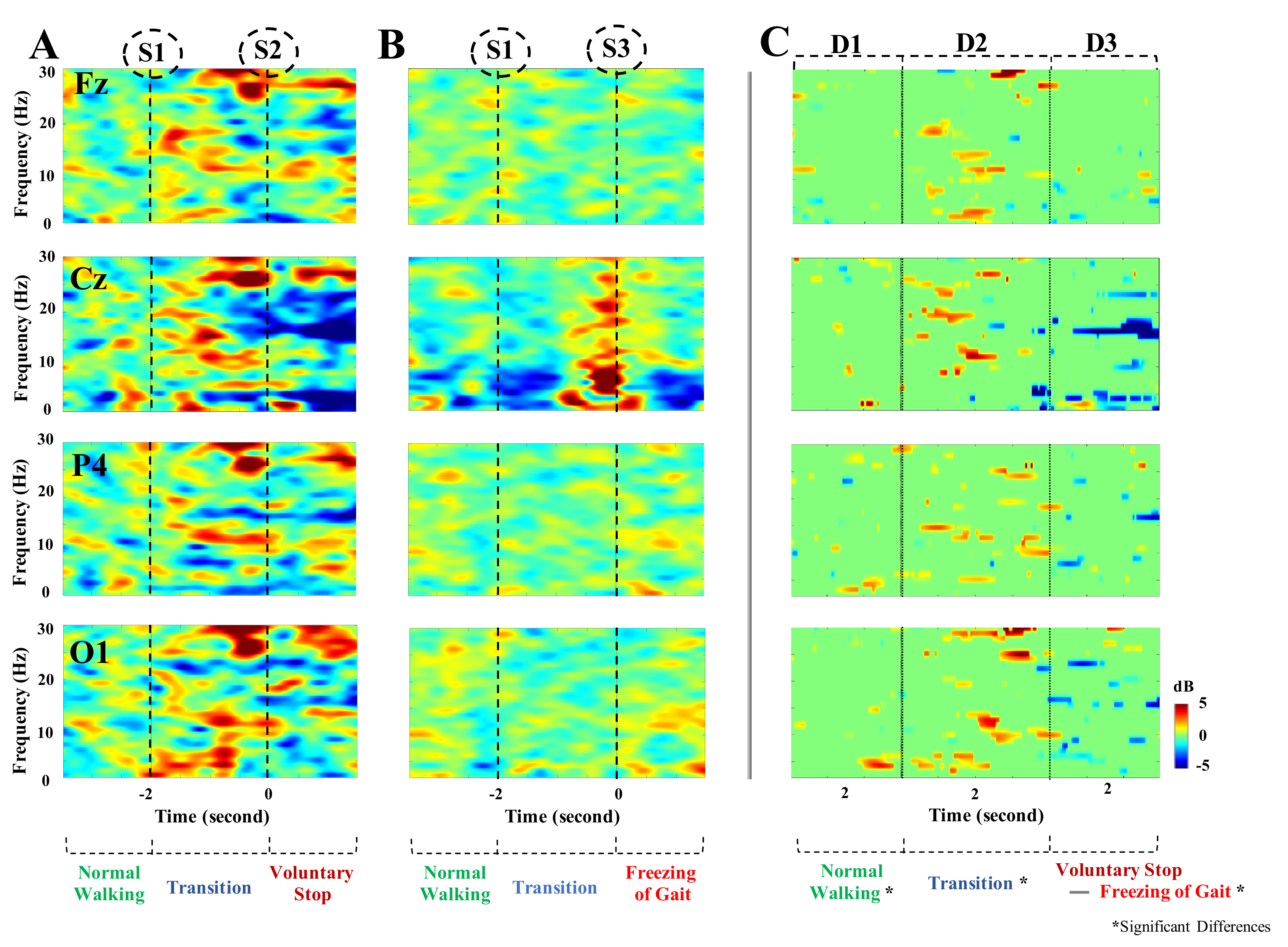}
\caption{Event-related band power dynamics between the voluntary stopping and FOG episodes during the TUG task. (A) The ERSP changes during the voluntary stopping episodes in the normal walking ($<$ -2 s), transition (-2 s to 0 s), and voluntary stopping ($>$ 0 s) periods at four electrodes (Fz, Cz, P4 and O1). (B) The ERSP changes during the FOG episodes in the normal walking ($<$ -2 s), transition (-2 s to 0 s), and FOG ($>$ 0 s) periods at four electrodes (Fz, Cz, P4 and O1). (C) Comparisons of the ERSP differences between the voluntary stopping and FOG episodes during the TUG task at four important electrodes (Fz, Cz, P4 and O1). S1 denotes the timing from normal walking to the transition period, S2 denotes the timing from the transition period to voluntary stop, S3 denotes timing from the transition period to FOG, D1 denotes difference during normal walking, D2 denotes the difference during the transition period, and D3 denotes the difference between the voluntary stopping and FOG episodes.}
    \label{fig5}
\end{figure*}

The two-dimensional images in Fig.~\ref{fig5} plot the ERSPs of the voluntary stopping (Fig.~\ref{fig5}A) and FOG episodes during the TUG task (Fig.~\ref{fig5}B) at the frontal, central, parietal, and occipital channels for patients with PD accompanied by FOG. In terms of the voluntary stopping episodes during the TUG task (Fig.~\ref{fig5}), a globally increased EEG power spectrum in the delta, theta, alpha and beta bands was observed in the transition period relative to normal walking. Furthermore, a partially decreased EEG power spectrum was observed during voluntary stopping relative to the transition period, and in particular, significantly reduced delta and beta power bands were observed during voluntary stopping at the Cz electrode.

We compared the ERSP differences between the FOG and voluntary stopping episodes while performing the TUG task during the normal walking, transition, and freezing/voluntary stopping periods at the four important electrodes (Fz, Cz, P4, and O1), as shown in Fig.~\ref{fig5}C. Our results showed significantly enhanced delta, theta, alpha and beta ERSPs in the voluntary stopping episodes relative to the FOG episodes during the normal walking and transition periods (p $<$ 0.05). The ERSP in the transition period preceding voluntary stopping demonstrated an increased power spectrum compared with that of the FOG episodes. More importantly, the ERSP during voluntary stopping showed a significantly decreased EEG power spectrum compared with that of the ERSP during freezing (p $<$ 0.05), and in particular, a distinctly reduced delta and low-beta EEG power spectrum was noted at the Cz electrode.

\begin{figure*}[!ht]
\center
    \includegraphics[width=\textwidth]{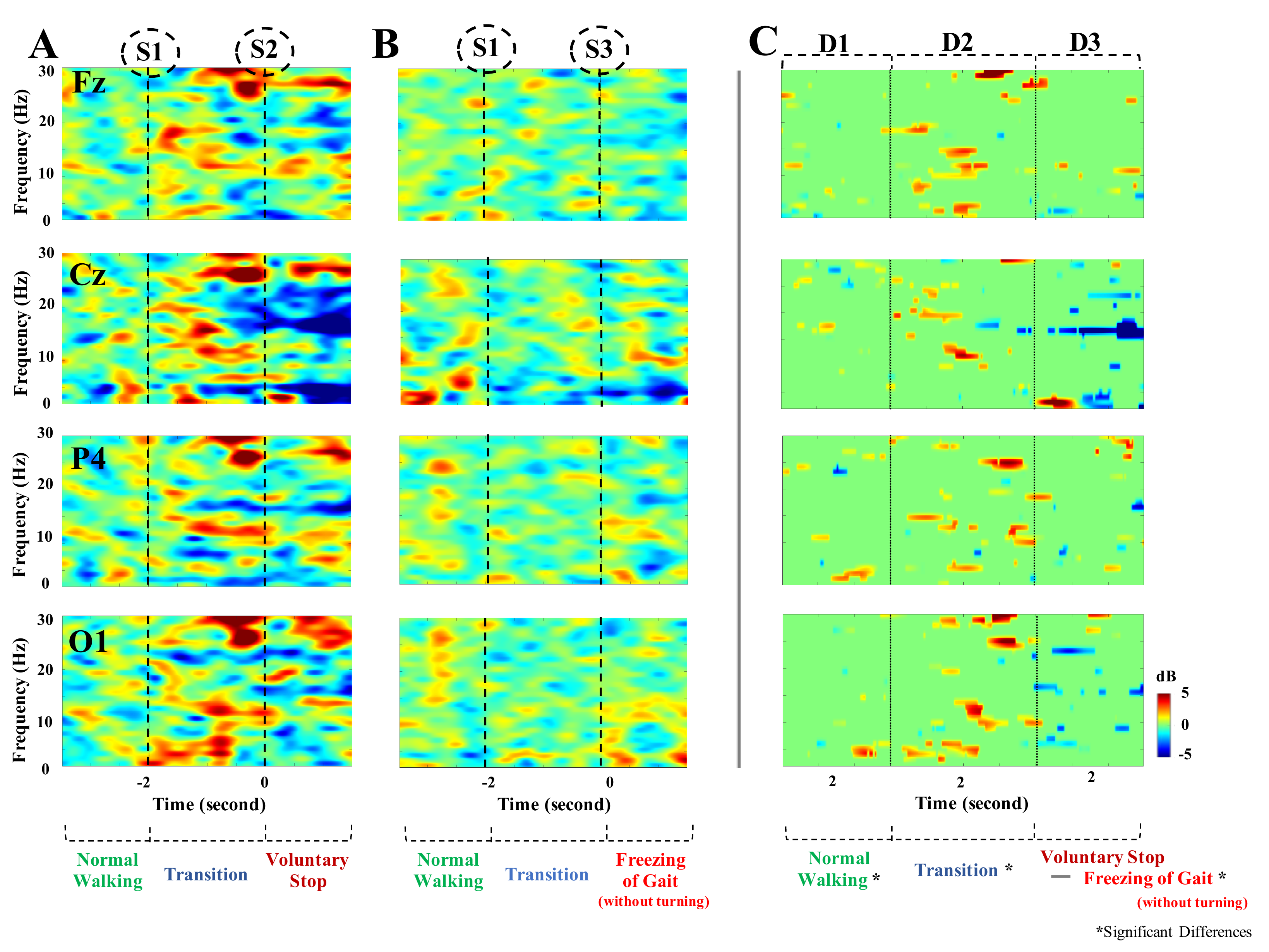}
\caption{Event-related band power dynamics between the voluntary stopping and FOG trials during the TUG task in which all the FOG episodes at turning were removed. (A) The ERSP changes during the voluntary stopping episodes in the normal walking ($<$ -2 s), transition (-2 s to 0 s), and voluntary stopping ($>$ 0 s) periods at four electrodes (Fz, Cz, P4 and O1). (B) The ERSP changes during the FOG in which all the FOG episodes at turning were removed in the normal walking ($<$ -2 s), transition (-2 s to 0 s), and FOG ($>$ 0 s) periods at four electrodes (Fz, Cz, P4 and O1). (C) Comparisons of the ERSP differences between the voluntary stopping and FOG in which all FOG episodes at turning were removed during the TUG task at four important electrodes (Fz, Cz, P4 and O1). S1 denotes the timing from normal walking to the transition period, S2 denotes the timing from the transition period to voluntary stop, S3 denotes timing from the transition period to FOG, D1 denotes difference during normal walking, D2 denotes the difference during the transition period and D3 denotes the difference between the voluntary stopping and FOG episodes.}
    \label{fig6}
\end{figure*}

To make the number of FOG trials and voluntary stop trials comparable, we studied the ERSPs of voluntary stopping and FOG trials after removing all the FOG episodes that occurred during turning. There were 119 FOG trials remaining after all the FOG trials at turning were removed. Fig. ~\ref{fig6} showed the ERSPs of voluntary stopping (Fig. ~\ref{fig6}A) and FOG (Fig. ~\ref{fig6}B) trials during the TUG task without the FOG during turning at the frontal, central, parietal, and occipital channels for the PD patients. The comparison between voluntary stopping trials and FOG trials without the turning FOG during normal walking, transition, and freezing of gait or voluntary stop periods at the Fz, Cz, P4 and O1 electrodes were shown in Fig. ~\ref{fig6}C. The observed results were similar to the ERSP results in which FOG trials at turning were also included. These results also showed an enhanced delta, theta, alpha and beta ERSP in voluntary stopping trials when compared to the ERSP in FOG trials during the normal walking and transition periods (p $<$ 0.05). ERSP in voluntary stopping trials showed a similarly increased power spectrum during the transition period in comparison to that of FOG trials. The results also showed a decreased power spectrum during a voluntary stop in comparison to the FOG (p $<$ 0.05). ERSP at the Cz electrode also showed a distinctly reduced delta and low-beta EEG power spectrum during the voluntary stop in contrast to that of freezing. 

\subsection{Brain hemisphere power dynamics between the voluntary stopping and FOG episodes}

\begin{figure*}[!t]
\centerline{\includegraphics[width=\textwidth]{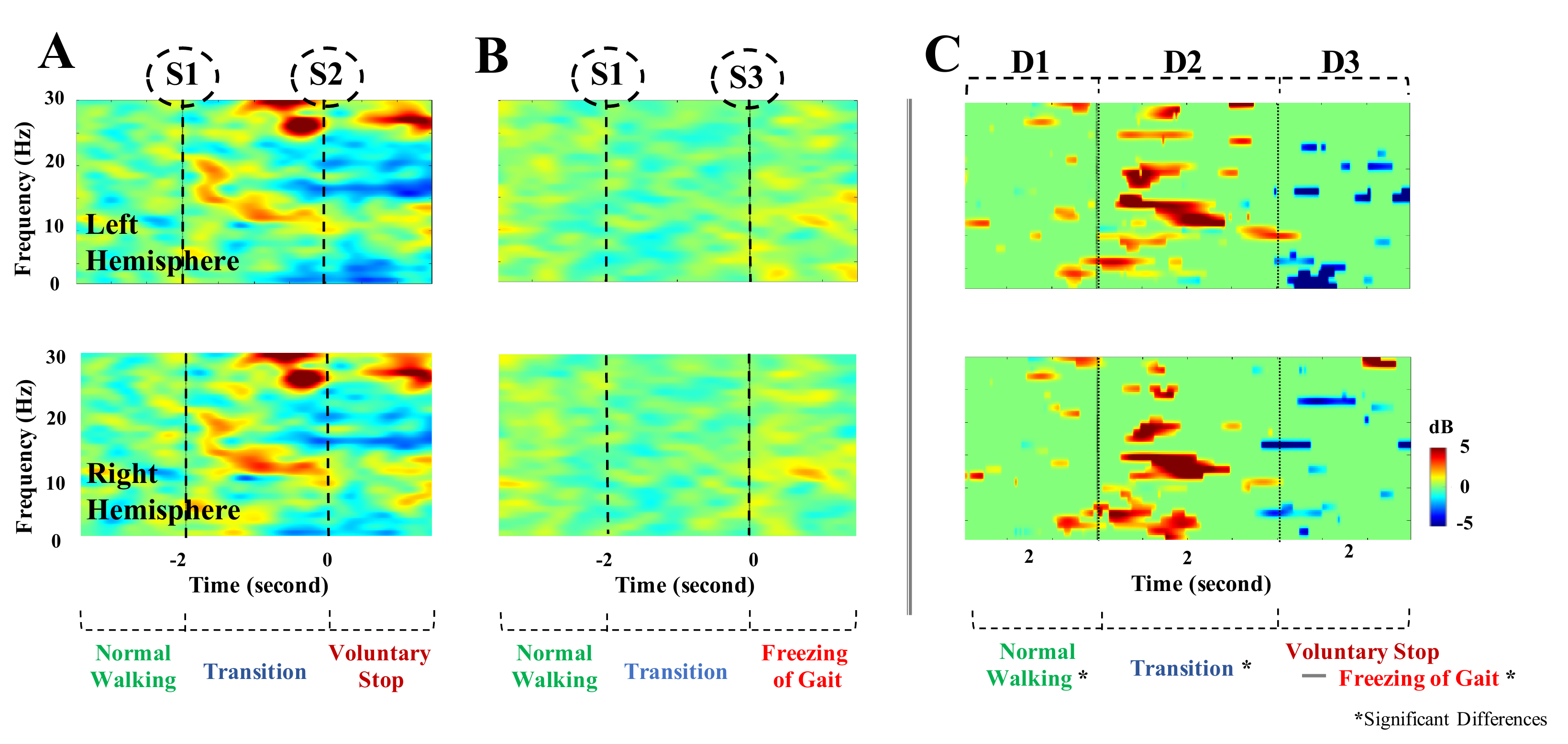}}
\caption{Brain hemisphere power dynamics during voluntary stopping and FOG episodes in the TUG task. (A) The ERSP changes during the voluntary stopping episodes in the normal walking ($<$ -2 s), transition (-2 s to 0 s), and voluntary stopping ($>$ 0 s) periods in the left and right hemispheres, respectively. (B) The ERSP changes during FOG episodes in the normal walking ($<$ -2 s), transition (-2 s to 0 s), and FOG ($>$ 0 s) episodes in the left hemisphere. (A3) Comparisons of the ERSP differences between the voluntary stopping and FOG episodes in the left and right hemispheres, respectively. (C) Comparisons of the ERSP differences between the voluntary stopping and FOG episodes in the left and right hemispheres, respectively.  S1 denotes the timing from normal walking to the transition period, S2 denotes the timing from the transition period to voluntary stop, S3 denotes timing from the transition period to FOG, D1 denotes difference during normal walking, D2 denotes the difference during the transition period, and D3 denotes the difference between the voluntary stopping and FOG episodes.}
\label{fig7}
\end{figure*}

Fig.~\ref{fig7} displays the results of the average EEG power spectra in the left and right hemispheres for the voluntary stopping and FOG episodes during the TUG task. For the voluntary stopping episodes during the TUG task, as shown in Figs.~\ref{fig7}A, a partial decrease in the delta power and an increase in the beta power were observed in the left and right hemispheres in the transition period relative to normal walking. Furthermore, the ERSP maintained decreased delta and low-beta power and increased high-beta power during a voluntary stopping in the left and right hemispheres. For FOG episodes during the TUG task, as shown in Fig.~\ref{fig7}B, no significant change in the ERSP was noted except for a slight increase in the alpha power during freezing.

The differences in the brain hemisphere power dynamics between the voluntary stopping and FOG episodes during the TUG tasks are shown in Fig.~\ref{fig7}C. Specifically, our results demonstrated that the EEG power spectrum during the transition was significantly higher in the beta band and slightly lower in the delta band during the voluntary stopping episodes than during the FOG episodes (p $<$ 0.05). Our results showed a significantly lower power in the delta to low-beta bands during the voluntary stopping period than during the FOG period (p $<$ 0.05). Moreover, we also noted a significantly higher high-beta power during the voluntary stopping period than during the FOG episode (p $<$ 0.05).

\section{DISCUSSION}
This study examined the ERSPs during voluntary stopping and freezing episodes using a revised TUG task with a voluntary stopping protocol. Our findings demonstrated the potential of ERSP analysis for identifying voluntary initiation and termination of the gait cycle. One key finding of this study was the increase in the EEG power in the central brain area before the onset of freezing episodes relative to that of normal walking. This finding resembles that of a recent study \cite{b7} that calculated the mean power spectral density, as shown in Table~\ref{tab1}. It is particularly interesting that both studies found an increase in the EEG beta power although different power measures and transition period definitions were applied.

In our study, the ERSP changes in all 178 FOG episodes were individually analysed to evaluate the statistical significance of the results. Each FOG episode showed a similar increase in power in the central brain area before the onset of freezing. Another key observation was that the ERSP changes for voluntary stopping episodes showed an elevated high-beta power at the Fz, Cz, P4, and O1 electrodes. These changes were verified in each voluntary stopping episode, and every episode showed a significantly increased high-beta power at these four electrodes. In addition, decreases in the EEG power were observed in the frontal, central, parietal and occipital areas during the voluntary stopping compared with the FOG episodes.

\subsection{FOG episodes related to EEG power dynamics in the central area}

Tracking and identifying FOG episodes by characterising their EEG signatures may provide a potential method for therapeutic prediction and alleviation of freezing episodes in patients with PD. Our study presented a novel finding that the transition from normal walking to freezing was associated with a significant increase in the EEG power in the central brain area. The results also showed an increase in delta and theta activity during the transition to freezing in the central area. The enhanced EEG power may reflect the brain's inhibitory processes activated by freezing episodes, mainly in the central area.

Furthermore, clinical studies have shown that freezing behaviour in PD is associated with a paroxysmal increase in theta oscillations  (5 - 7 Hz), known as ``trembling in place'' \cite{b28}. The changes observed in the EEG power spectrum in the theta band during freezing may be due to mechanical oscillations transmitted to the scalp electrodes. It is worth noting that the increase in the theta power was unlikely due to global motor interference, as we observed a significant increase in the theta power only in the central area. Perhaps a more acceptable explanation is that a dysfunctional neuronal circuit in subcortical brain structures drives the creation of theta oscillations in patients with PD \cite{b29}. A similar mechanism was recently found to underlie the oscillations observed during FOG \cite{b30}.

The results also showed that the beta activity in the central area was enhanced during this transition period. The change in the beta power might be associated with the facilitation of postural activities, including a tonic holding contraction and inhibition of voluntary movement \cite{b31}. Moreover, beta suppression is critical for the facilitation of continuous movement sequences \cite{b32}, suggesting that high-beta activity may interface with anticipatory postural adjustments in preparation for stepping, which can facilitate excessive postural contraction of the lower limbs associated with FOG episodes \cite{b33}.

\subsection{Acute prediction and detection of FOG episodes}

In previous studies, the transition period prior to FOG was defined as the time window of 5 to 1 s before the occurrence of freezing \cite{b7, b21}. However, this study analysed a shorter transition period of 2 s before freezing, which might have more clinical utility, as gait parameters usually change prior to freezing within a short time window. Compared to those of previous studies, the EEG power in this study showed similar changes at the Cz electrode but different changes at the O1 electrode in the transition period from normal walking, suggesting that the EEG in the occipital area may be susceptible to the preparation of freezing. 
In this study, considering a 2-s transition period to a FOG event, we observed an increase in the beta power in the central region of the brain, as with the 5-s transition period, compared to normal walking. 
However, we also observed an increase in the alpha power in the central region, while an increase in the theta power was observed during the 5-s transition period compared to the normal walking period. When the 5-s transition period was compared to the normal walking period, increases in the alpha and beta power were observed in the parietal area; however, only a partial increase in the beta power was observed when the 2-s transition period was compared to the normal walking period. While an increase in the beta power in the occipital area was observed only when the 5-s transition period was compared to normal walking, we observed increases in both the alpha and beta power at the occipital region when comparing the 2-s transition period to normal walking.

\subsection{EEG power dynamics between involuntary and voluntary stopping}

Currently, few studies \cite{b7, b9} have highlighted the distinctions of different motor inhibition processes, and thus, investigating brain dynamics under conditions of involuntary stopping (freezing) and voluntary stopping is insightful. Our results showed that the voluntary stopping period had a significantly decreased EEG power spectrum compared to that in the freezing period, and a distinctly different pattern in the delta and low-beta power was particularly observed in the central area. These suppressions may reflect motion preparation and execution \cite{b34} as well as EEG oscillatory patterns, reflecting the process of inhibitory control in the brain \cite{b35, b36}, which controls the cessation of foot movement when patients hear ``stop'' cues during TUG trials.

In fact, the voluntary stopping task in our TUG task with the voluntary stopping component is similar to the stop-signal paradigm, wherein a participant engaged in a task stops their primary task when presented with a signal to stop \cite{b11}. As with the stop-signal paradigm, voluntary stopping during the TUG task may also involve initiation and inhibition. The transition period in the voluntary stopping episode is similar to response inhibition, where the subject is required to stop walking after initiating movements quickly. In terms of the transition period, our results showed increased EEG power in the voluntary stopping episodes compared to the FOG episodes. This strengthening of the EEG power indicates an effective preparation stage for response inhibition \cite{b29}, whereas a low EEG power spectrum during freezing may involve incomplete or ineffective inhibitory control over some movement options.

\begin{table}
\caption{EEG dynamics during the transition to FOG }
\label{table}
\setlength{\tabcolsep}{3pt}
\begin{tabular}{|p{50pt}|p{35pt}|p{40pt}|p{40pt}|p{40pt}|}
\hline
\multirow{2}{*}{EEG Pattern}& 
\multirow{2}{*}{Frequency}& 
\multicolumn{3}{c|}{Observation at Brain Cortices} \\
\cline{3-5}
& & Central & Parietal & Occipital \\
\hline
Power spectral & $\delta$ & N/A & N/A & N/A \\
\cline{2-5}
density (PSD) & $\theta$ & $\uparrow$ & N/A & N/A \\
\cline{2-5}
with a 5-s & $\alpha$ & N/A & $\downarrow$ & N/A \\
\cline{2-5}
transition \cite{b7}$^{\mathrm{a}}$ & $\beta$ & $\uparrow$ & $\uparrow$ & $\uparrow$ \\
\hline
ERSP with a  & $\delta$ & N/A & N/A & N/A \\
\cline{2-5}
2-s  & $\theta$ & N/A & N/A & N/A \\
\cline{2-5}
transition (Our & $\alpha$ & $\uparrow$ & N/A & $\uparrow$ \\
\cline{2-5}
findings)$^{\mathrm{b}}$ & $\beta$ & $\uparrow$ & $\uparrow$ & $\uparrow$ \\
\hline
\multicolumn{5}{|c|}{$^{\mathrm{a}}$ The transition period, defined as 5 s before a FOG episode.}\\
\multicolumn{5}{|c|}{The extracted period of normal walking lasts 5 s.}\\
\multicolumn{5}{|c|}{$^{\mathrm{b}}$ The transition period, defined as 2 s before a FOG episode.}\\
\multicolumn{5}{|c|}{The extracted period of normal walking lasts 2 s.}\\
\multicolumn{5}{|c|}{ $\uparrow$ $\downarrow$ The up and down arrows represent increased and decreased}\\
\multicolumn{5}{|c|}{EEG power in the transition period compared}\\
\multicolumn{5}{|c|}{with the normal walking period, respectively.}\\
\hline

\end{tabular}
\label{tab1}
\end{table}

\section{Conclusion}

This study investigated the brain dynamics of FOG and voluntary stopping episodes using EEG data collected from patients with PD accompanied by FOG. Our findings highlighted that FOG episodes were associated with abnormal EEG dynamics and that voluntary stopping could be discriminated from FOG episodes. Comparing the transition to freezing period to the freezing period itself, our findings show that freezing episodes are associated with significantly increased theta and alpha band power within the central and occipital areas. Furthermore, the EEG power significantly decreased during the voluntary stopping period compared with the FOG period. Our results provide novel insights into the rapid transition dynamics underlying the phenomenon of FOG and may provide a potential means for the therapeutic prediction and alleviation of freezing episodes in susceptible patients. These findings are very useful for the development of future technologies that predict FOG episodes, as they suggest that voluntary stopping will not activate false positives, allowing for the accurate detection of freezing.

\ifCLASSOPTIONcaptionsoff
\newpage
\fi



%

\end{document}